\documentclass[aps,floatfix,preprint,showpacs,preprintnumbers,nofootinbib,superscriptaddress,natbib]{revtex4}

\usepackage{graphicx,float}
\usepackage{epsfig}		
\usepackage{dcolumn}
\usepackage{bm}
\usepackage{subfigure}

\usepackage[all]{xy}
\usepackage{amsmath,upgreek}
\usepackage{amssymb}

\usepackage{pdfpages}
\usepackage{color}
\usepackage{graphicx,epstopdf}
\usepackage[colorlinks,hyperindex]{hyperref}

\def\0{\mbox{\tiny $0$}}
\def\1{\mbox{\tiny $1$}}
\def\2{\mbox{\tiny $2$}}
\def\3{\mbox{\tiny $3$}}
\def\4{\mbox{\tiny $4$}}
\def\5{\mbox{\tiny $5$}}
\def\6{\mbox{\tiny $6$}}
\def\7{\mbox{\tiny $7$}}
\def\8{\mbox{\tiny $8$}}
\def\9{\mbox{\tiny $9$}}

\def\f14{\mbox{\tiny $\frac{1}{4}$}}

\begin{document}

\title{On the elementary information content of thermodynamic ensembles}
\author{Alex E. Bernardini}
\email{alexeb@ufscar.br}
\affiliation{~Departamento de F\'{\i}sica, Universidade Federal de S\~ao Carlos, PO Box 676, 13565-905, S\~ao Carlos, SP, Brasil.}

\begin{abstract}
Extending the definition of phase-space (Wigner) quantum projectors to thermodynamic ensembles usually results into an efficient platform for quantifying their elementary information content.
Given the spectral decomposition profile of a thermalized quantum system, general expressions for the quantum purity quantifier, $\mathcal{P}(\beta)$, and for phase-space projectors related to the quantum fidelity, $\mathcal{F}(\beta)$, are explicitly derived in terms of an explicit correspondence with the related partition function, $\mathcal{Z}(\beta)$.
Besides quantifying the storage of information capacity of thermodynamical ensembles, the tools here introduced extend the role of the partition function in expressing the quantum behavior of thermodynamic ensembles.
\end{abstract}

\date{\today}
\maketitle

\paragraph*{Introduction --} Even considering that the concept of statistical thermodynamic equilibrium described by the phase-space Gibbs-Boltzmann formula can not be trivially generalized to quantum mechanics, the Weyl-Wigner phase-space formalism \cite{Case,Wigner} sufficient and solid grounds for the inception of the quantum statistical mechanics \cite{Hillery}.
As a consequence, the thermodynamics of quantum ensembles embedded into the phase-space quantum mechanical description of Nature exhibits an enlarged set of encoded information which may (or not) be summarized by the partition function of a thermalized system.

Thermal average values obtained for canonical ensembles as weighted sums over all (quantum) states, once resumed by its statistical partition function, are operationally useful insofar as one can quantum mechanically calculate the Hamiltonian system energy levels and their related properties.
In this context, besides the well-stablished quantifiers for the storage of energy \cite{Novo01} -- namely related to classical and quantum results for internal energy and heat capacity -- general expressions for storage of information quantifiers: the quantum purity, $\mathcal{P}(\beta)$, and phase-space projectors related to the quantum fidelity, $\mathcal{F}(\beta)$, can also be explicitly obtained in terms of the quantum system related partition function, $\mathcal{Z}(\beta)$.
To demonstrate such an assertion, canonical ensembles constructed as thermalized statistical mixtures supported by the Weyl-Wigner formalism are considered. Their construction aspects straightforwardly provide the results for measures of quantum purity, through which temperature dependent decoherence effects can be explicitly quantified in terms of the ensemble partition function.

Besides quantifying the storage of information capacity, our results suggest that the quantum projector tools obtained from the phase-space description of thermodynamic ensembles extend the role of the partition function in expressing the system quantum behavior.

\paragraph*{Quantum Projectors for Thermodynamic Ensembles --} Departing from the one-dimension Weyl transform \cite{Case} of a quantum operator, $\hat{\rho}$,
\begin{equation}
\rho^W(q, p)
= 2\hspace{-.1cm} \int^{+\infty}_{-\infty} \hspace{-.4cm}du\,\exp{\left(2\,i \,p\, u/\hbar\right)}\,\langle q - u | \hat{\rho} | q + u \rangle=2\hspace{-.1cm} \int^{+\infty}_{-\infty} \hspace{-.4cm} dv \,\exp{\left(-2\, i \,q\, v/\hbar\right)}\,\langle p - v | \hat{\rho} | p + v\rangle,
\end{equation}
where $q$ and $p$ are position and momentum coordinates, the Wigner function for a quantum state, $W(q, p)$, is expressed by $\rho^W(q, p)$, when $\hat{\rho}$ is identified with the quantum mechanical density operator.
For $\hat{\rho}$ written in terms of quantum states as $\hat{\rho} =|\phi \rangle \langle \phi |$, one then has
\begin{equation}
 h^{-1} \hat{\rho} \to W(q, p) = (\pi\hbar)^{-1} 
\int^{+\infty}_{-\infty} \hspace{-.3cm}du\,\exp{\left(2\, i \, p \,u/\hbar\right)}\,
\phi^{\ast}(q - u)\,\phi(q + u),
\end{equation}
which exhibits the properties of a real-valued {\em quasi}-probability distribution and has a number of subtle properties connected to the matrix operator quantum mechanics.
For instance, the trace of the product between $\hat{\rho}$ and a generic operator, $\hat{O}$, results into average values, $\langle O \rangle$, evaluated by the product of their Weyl transforms integrated over the phase-space plane, $q-p$, as \cite{Wigner,Case}
\begin{equation}
Tr_{\{q,p\}}\left[\hat{\rho}\hat{O}\right] \to \langle O \rangle = 
\int^{+\infty}_{-\infty} \hspace{-.3cm}{dp}\int^{+\infty}_{-\infty} \hspace{-.3cm} {dq}\,W(q, p)\,{O^W}(q, p).
\label{five}
\end{equation}
Of course, it is constrained by the normalization condition of $\hat{\rho}$, $Tr_{\{q,p\}}[\hat{\rho}]=1$, which suggests the preliminary interpretation of $W(q, p)$ as a probability distribution.
In particular, such trace analog operations result into the projection property described by
\begin{equation}\label{qp}
\mathcal{F}^{ab} = Tr_{\{q,p\}}\left[\hat{\rho}^a\hat{\varrho}^b\right] = 2\pi\hbar
\int^{+\infty}_{-\infty} \hspace{-.3cm}{dp}\int^{+\infty}_{-\infty} \hspace{-.3cm} {dq}\,W^a(q, p)\,\mathcal{W}^b(q, p) = \vert\langle \phi\vert\varphi\rangle\vert^2,
\end{equation}
for $\hat{\rho}^a =|\phi \rangle \langle \phi |$ identified with $W^a(q, p)$, and $\hat{\varrho}^b =|\varphi \rangle \langle \varphi|$ identified with $\mathcal{W}^b(q, p)$. From Eq.~(\ref{qp}), the quantum purity, $\mathcal{P}$, is also obtained from a straightforward trace operation given by
\begin{equation}
\mathcal{P} = Tr_{\{q,p\}}[\hat{\rho}^2] = 2\pi\hbar\int^{+\infty}_{-\infty} \hspace{-.3cm}{dp}\int^{+\infty}_{-\infty} \hspace{-.3cm} {dq}\,W(q, p)^2.
\label{pureza}
\end{equation}
For gaussian states, the square root of $\mathcal{F}^{ab}$ corresponds to the so-called quantum fidelity between $\hat{\rho}^a$ and $\hat{\varrho}^b$, and for identical pure states, with $\hat{\rho}^a \equiv \hat{\varrho}^b$, $\mathcal{F}^{ab}$ is reduced to $\mathcal{P} = Tr_{\{q,p\}}[\hat{\rho}^2] = 1$. 
Likewise, as it shall be demonstrated in the following, such information quantifiers statistically related to the properties of the density matrix can also be computed for canonical ensembles, in terms of their associated partition functions.

From this point, as a matter of convenience, the above framework shall be reconfigured in terms of a dimensionless form of the Wigner function, $\mathcal{W}(x, \, k;\,\tau)$, written in terms of dimensionless variables, $x = \left(m\,\omega\,\hbar^{-1}\right)^{1/2} q$ and $k = \left(m\,\omega\,\hbar\right)^{-1/2}p$, for which mass and energy scales, $m$ and $\hbar \omega$, have been provided, and a dimensionless time-like quantity, $\tau = \omega t$, has been described in terms of an arbitrary angular frequency, $\omega$.
It is resumed by \cite{NovoPaper}\footnote{The correspondence between $\psi(x,\,\tau)$ and $\phi(q;\,t)$ is consistent with their normalization constraints,
\begin{equation}
\int^{+\infty}_{-\infty} \hspace{-.2 cm}{dx}\,\vert\psi(x;\,\tau)\vert^2 =\int^{+\infty}_{-\infty}\hspace{-.2 cm}{dq}\,\vert\phi(q;\,t)\vert^2 = 1.
\end{equation}}
\begin{eqnarray}
\mathcal{W}(x, \, k;\,\tau) &=& \pi^{-1} \hspace{-.3cm}\int^{+\infty}_{-\infty} \hspace{-.3cm}dy\,\exp{\left(2\, i \, k \,y\right)}\,\psi(x - y;\,\tau)\,\psi^{\ast}(x + y;\,\tau),\,\,\,\,
\end{eqnarray}
with $y = \left(m\,\omega\,\hbar^{-1}\right)^{1/2} u$, and where $\hbar$ has been absorbed by the (implicit) phase-space volume integration.

From now on, considering that the quantum propagator (Green's function) for a time-independent dimensionless Hamiltonian $\mathcal{H}$ could be expressed by
\begin{equation}
\Delta(x,t;\,x^{\prime},0) = \langle x \vert \exp(-i\,\tau\, \hat{\mathcal{H}}) \vert x^{\prime} \rangle,
\end{equation}
a thermal density matrix thermodynamically engendered by $\mathcal{H}$ can be derived from $\Delta(x,\tau;\,x^{\prime},0)$.
For a canonical ensemble in equilibrium with a heat reservoir at temperature $T$, the associated thermal density matrix, $\rho(x,\,x^{\prime};\,\beta \hbar \omega)$, can be obtained \cite{Hillery,Ballentine} by replacing the above related time-dependence, $\tau$, by $- i \, \beta \hbar \omega$, where $\beta = 1/ k_{B} T$, and $k_{B}$ is the Boltzmann constant.
In this case, one has
\begin{equation}
\rho(x,\,x^{\prime};\,\beta \hbar \omega) = \Delta(x,- i \, \beta;\,x^{\prime},0) = \sum_n \exp(-\beta\hbar\omega\, \varepsilon_n)\,\psi^{*}_n({x})\psi_n({x^{\prime}}),
\end{equation}
where quantum eigenstates and eigenenergies, $\psi_n({x})$ and $\varepsilon_n$, are implicitly constrained by the eigenvalue equation $\mathcal{H}\,\psi_n({x}) = \varepsilon_n\,\psi_n({x})$.

In the coordinate representation of the state operator for the canonical ensemble, the functional $\rho(x,\,x^{\prime};\,\beta \hbar \omega)$ can have its delocalization aspects parameterized by displacement relations, $x \to x+y$ and $x^{\prime} \to x - y$. It leads to the $y$-Fourier transform of $\rho(x+y,\,x-y;\,\beta \hbar \omega)$ identified with a thermalized phase-space probability distribution, $\Omega(x, \, k;\,\beta)$, as
\begin{eqnarray}
\label{W222}
\Omega(x, \, k;\,\beta) &=&  \pi^{-1} \int^{+\infty}_{-\infty} \hspace{-.3cm}dy\,\exp{\left(2\, i \, k \,y\right)}\,\rho(x+y,\, x-y;\,\beta\hbar\omega),
\end{eqnarray}
which satisfies the Bloch equation \cite{Bloch,Hillery}, 
\begin{equation}
\frac{\partial {\Omega}}{\partial \beta} = - \hat{\mathcal{H}}{\Omega} = -{\Omega}\hat{\mathcal{H}},
\end{equation}
with ${\Omega}(\beta=0)$ working as an identity operator.

From Eq.~(\ref{W222}), one thus identifies the correspondence between $\rho(x+y,\, x-y;\,\beta\hbar\omega)$ and the canonical ensemble partition function, $\mathcal{Z}(\beta)$, by means of a trace operation explicitly given by
\begin{equation}\label{stand}
\mathcal{Z}(\beta) = Tr\left[\exp(-\beta\hbar\omega\, \hat{\mathcal{H}})\right] = \int^{+\infty}_{-\infty} \hspace{-.3cm} dx\,\int^{+\infty}_{-\infty} \hspace{-.3cm}dk\,\,\Omega(x, \, k;\,\beta) = \sum_{n=0}^{\infty}\exp\left(-\beta\hbar\omega\,\varepsilon_n\right),
\end{equation}
from which a large set of systematic thermodynamic and statistical results can be evinced.
From Eq.~\eqref{W222}, the corresponding normalized Wigner function is then written as
\begin{eqnarray}
\label{W222B}
\mathcal{W}_\Omega(x, \, k;\,\beta) &=& \left(\pi\,\mathcal{Z}(\beta)\right)^{-1} \int^{+\infty}_{-\infty} \hspace{-.3cm}dy\,\exp{\left(2\, i \, k \,y\right)}\,\rho(x+y,\, x-y;\,\beta),
\end{eqnarray}
which is converted into the essential tool for defining quantum projectors for thermodynamic ensembles.

To clear up this point, one can firstly consider the ensemble of quantum states of a thermodynamical system $a$ described in terms of Wigner distributions written as
\begin{eqnarray}
\label{W222BBB}
\mathcal{W}^a_\Omega(x, \, k;\,\beta) &=&\frac{1}{\mathcal{Z}^a(\beta)}\sum_{n=0}^{\infty} a_n(\beta)\,\mathcal{W}^a_n(x, \, k),\end{eqnarray}
with $a_n(\beta) = \exp(-\beta\hbar\omega\,\varepsilon^a_n)$ and
\begin{eqnarray}
\mathcal{W}^a_n(x, \, k) &=& \pi^{-1} \int^{+\infty}_{-\infty} \hspace{-.3cm}dy\,\exp{\left(2\, i \, k \,y\right)}\,\psi^a_n(x - y)\,\psi^{a\ast}_n(x + y),
\end{eqnarray}
for the set of ortonormalized stationary states $\{\psi^a_n\}$.
Analogous definitions for a system $b$, with $b_n(\beta) = \exp(-\beta\hbar\omega\,\varepsilon^b_n)$, are also introduced.
In this case, the quantum projector from Eq.~(\ref{qp}) leads to
\begin{equation}\label{qp2}
\mathcal{F}^{ab} \to \mathcal{F}^{ab}(\beta,\,\beta') = \frac{1}{\mathcal{Z}^a(\beta)\,\mathcal{Z}^b(\beta')}\sum_{n,\ell = 0}^{\infty} a_n(\beta)\,b_{\ell}(\beta')
\vert \alpha_{n\ell}\vert^2, \qquad \mbox{with}\,\,\, \alpha_{n\ell}=\langle \psi^a_n\vert\psi^b_{\ell}\rangle.
\end{equation}
By observing that $a_n(0) = b_{\ell}(0) = 1$, as well as $$\sum_{\ell=0}^{\infty} \vert \alpha_{n\ell}\vert^2 = \sum_{n=0}^{\infty} \vert \alpha_{n\ell}\vert^2 = \sum_{n=0}^{\infty}\langle \psi^b_{\ell}\vert\psi^a_n\rangle\langle \psi^a_n\vert\psi^b_{\ell}\rangle = \langle \psi^b_{\ell}\vert\left(\sum_{n=0}^{\infty}\vert\psi^a_n\rangle\langle \psi^a_n\vert\right)\vert \psi^b_{\ell}\rangle = \langle \psi^b_{\ell}\vert\\\mathbb{I} \vert \psi^b_{\ell}\rangle  =1,$$
one notices that $\mathcal{F}^{ab}(0,\,\beta') = {1}/{\mathcal{Z}^a(\beta\to 0)} = \mathcal{F}^{ab}(\beta,\,0) = {1}/{\mathcal{Z}^a(\beta'\to 0)} = \mathcal{F}^{ab}(0,\,0) = 0$, which corresponds to the lower bound value of $\mathcal{F}^{ab}(\beta,\,\beta')$.
More relevantly, for $a \equiv b$, one has $\alpha_{n\ell} = \delta_{n\ell}$ and, therefore,
\begin{equation}\label{qpqp}
\mathcal{F}^{aa}(\beta,\,\beta') = \frac{\mathcal{Z}^a(\beta+\beta')}{\mathcal{Z}^a(\beta)\,\mathcal{Z}^a(\beta')},
\end{equation}
which corresponds to the quantum fidelity between the same canonical ensembles at different temperatures related to $\beta$ and $\beta'$. It explicitly yields the quantum purity of the thermodynamic ensemble $a$ in terms of $\mathcal{Z}^a$ as
\begin{equation}\label{qpqpqp}
\mathcal{P}^{a}(\beta) = \mathcal{F}^{aa}(\beta,\,\beta) = \frac{\mathcal{Z}^a(2\beta)}{\left(\mathcal{Z}^a(\beta)\right)^2}.
\end{equation}
As prescribed by the analytical property of the quantum fidelity, the result from Eq.~(\ref{qpqp}) simply allows for quantitatively confronting the storage information capacity (related to quantum purity) between identical quantum ensembles at different thermodynamic regimes. If asymptotic temperature regimes, with $T\to \infty$, are admitted as the classical limit, the fidelity between classical and quantum thermodynamic ensembles is null.

At this point, it is also relevant to notice that if a system is subdivided into $N$ non-interacting distinguishable sub-systems, $a_1$, $a_2$, $\dots$, $a_N$, then the partition function of the global system is expressed by the product of the individual partition functions, $\mathcal{Z}^a = \prod_{s=1}^N \mathcal{Z}^{a_s}$, which straightforwardly yields
\begin{equation}\label{qpqpqpSA}
\mathcal{P}^{a}(\beta) = \frac{\mathcal{Z}^a(2\beta)}{\left(\mathcal{Z}^a(\beta)\right)^2} = \prod_{s=1}^N \frac{\mathcal{Z}^{a_s}(2\beta)}{\left(\mathcal{Z}^{a_s}(\beta)\right)^2} =  \prod_{s=1}^N  \mathcal{P}^{a_s}(\beta),
\end{equation}
which shows that the quantum purity, $\mathcal{P}$, exhibits subaditivity properties similar to those ones of the partition function, $\mathcal{Z}$.
Likewise, if the sub-systems are quantum mechanically identical one to each other, i.e. $a_1 \equiv a_2 \equiv \dots \equiv a_N = A $, the total partition function must be divided by a factor $N!$ so as to ensure the right number of microstates.
It results into $\mathcal{Z}^a = (\mathcal{Z}^{A})^N/N!$ which yields $\mathcal{P}^a = (\mathcal{P}^{A})^N/N!$, as expected.

Finally, given the above correspondence with the partition function, a storage of information capacity related to $\mathcal{P}(\beta)$ can be introduced by the dimensionless quantities,
\begin{equation}\label{qua16HObb}
\epsilon^{\mathcal{P}}(\beta) = 
\beta \frac{\partial}{\partial \beta} \ln\left(\mathcal{P}(\beta)\right),
\end{equation}
and
\begin{equation}\label{qua16HO}
\mathcal{C}^{\mathcal{P}}(\beta) = - \beta^2 \frac{\partial^2}{\partial \beta^2} \ln\left(\mathcal{P}(\beta)\right),
\end{equation}
which are respectively analogous to the internal energy and to the heat capacity of a thermodynamic ensemble, when $\mathcal{P}(\beta)$ is replaced by $1/\mathcal{Z}(\beta)$.
As it shall be discussed in the following, both quantities define {\em plateaus} of information which can be related to some temperature dependent storage capacity as well as it is noticed from the energy-like quantities defined in terms of $\ln(\mathcal{Z}(\beta))$ derivatives in statistical thermodynamics. Apart from the nomenclature analogy, as well known, the thermodynamic heat capacity, $\mathcal{C}(\beta)$, expresses how much energy one needs to change the temperature of a given mass. It reflects into its capacity of storing heat. In the case of the replacement of $\mathcal{Z}(\beta)$ by $1/\mathcal{P}(\beta)$, $\epsilon^{\mathcal{P}}(\beta)$ (complemented by $\mathcal{C}^{\mathcal{P}}(\beta)$) expresses the rate(s) of change of stored information which is gained(lost) throughout the ensemble evolution from statistical mixtures (pure states) to pure states (statistical mixtures) --  they quantify the quantum coherence behavior in terms of the temperature associated parameter, $\beta$.

\paragraph*{Harmonic Oscillator --} A natural {\em test platform} for the above discussed quantities is the quantum harmonic oscillator (HO) driven by the dimensionless Hamiltonian equation, 
\begin{equation}\label{qua16HO3}
\mathcal{H}^{HO} \phi^{HO}_n(x) = \frac{1}{2}\left(k^2+ x^{2} \right) \phi^{HO}_n(x) = (n + 1/2)\,\phi^{HO}_n({x}),
\end{equation}
with Wigner functions associated to $\phi^{HO}_n({x})$ written as
\begin{equation}\label{qua16HO2}
\mathcal{W}^{HO}_n(x, \, k) = (-1)^n \pi^{-1} \exp[-(k^2+ x^{2})]\,\mathcal{L}_n[2(k^2+ x^{2})],
\end{equation}
where $\mathcal{L}_n$ are the {\em Laguerre polynomials}, which lead to the thermalized Wigner function (cf. Eq.~(\ref{W222BBB})) written as \cite{Hillery}
\begin{eqnarray}
\label{W222BHO}
\mathcal{W}^{HO}_\Omega(x, \, k;\,\beta) &=&
\pi^{-1}
\tanh\left(\beta\hbar\omega/2\right)
\exp\left[-\tanh\left(\beta\hbar\omega/2\right)
(k^2+ x^{2})\right],
\end{eqnarray}
which results into
\begin{equation}\label{qpqpqpHO}
\mathcal{Z}^{HO}(\beta) = \frac{1}{2\sinh\left(\beta\hbar\omega/2\right)},
\end{equation}
and, consequently,
\begin{equation}\label{qpqpqpHOHO}
\mathcal{F}^{HO}(\beta,\,\beta') = \frac{2}{\coth\left(\beta\hbar\omega/2\right)+\coth\left(\beta'\hbar\omega/2\right)}\quad\mbox{and}\quad
\mathcal{P}^{HO}(\beta) = \tanh\left(\beta\hbar\omega/2\right),
\end{equation}
respectively for quantum fidelity and quantum purity.

\begin{figure}[b!]
\vspace{-.5cm}\includegraphics[scale=0.4]{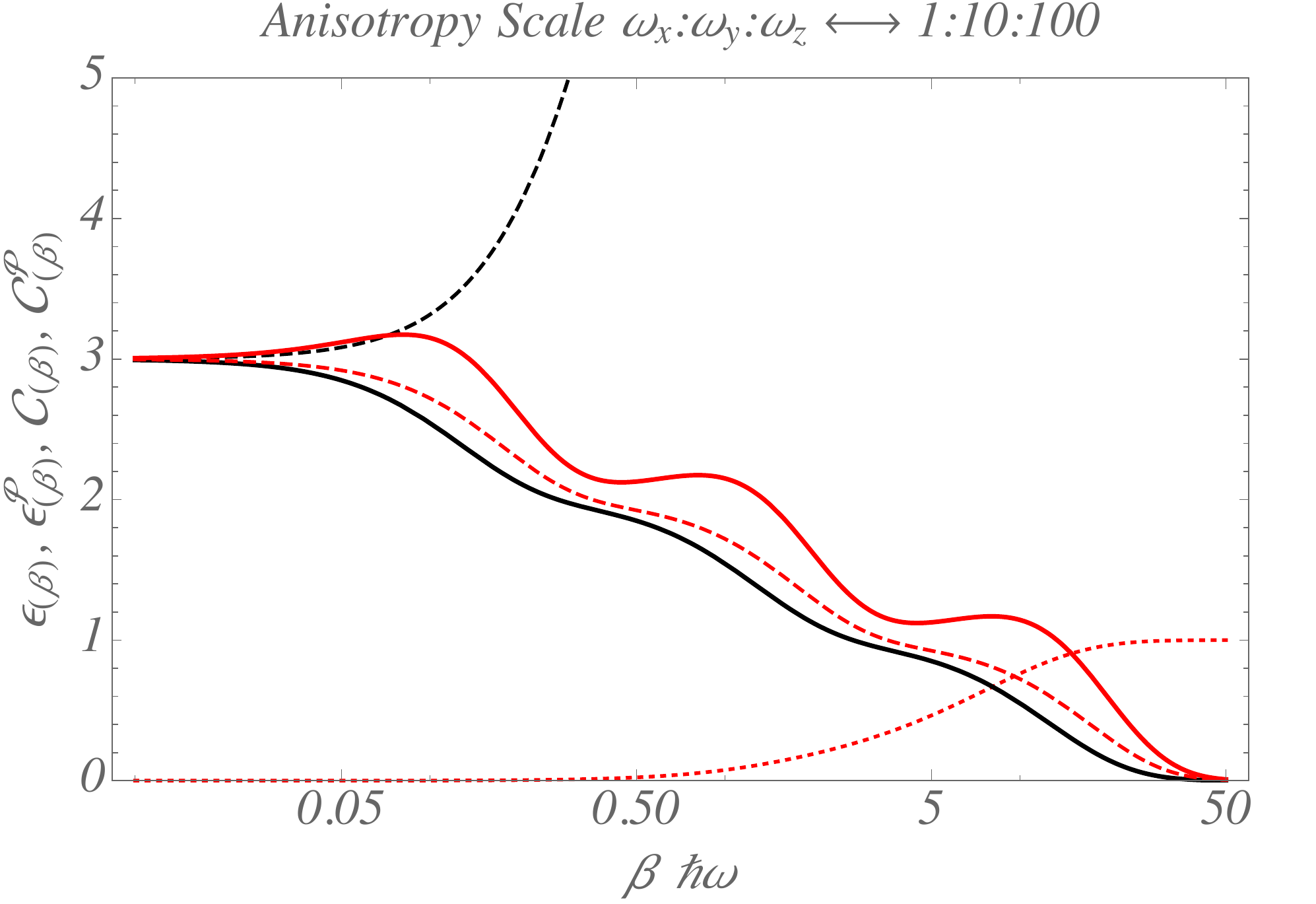}
\vspace{-.5cm}\includegraphics[scale=0.4]{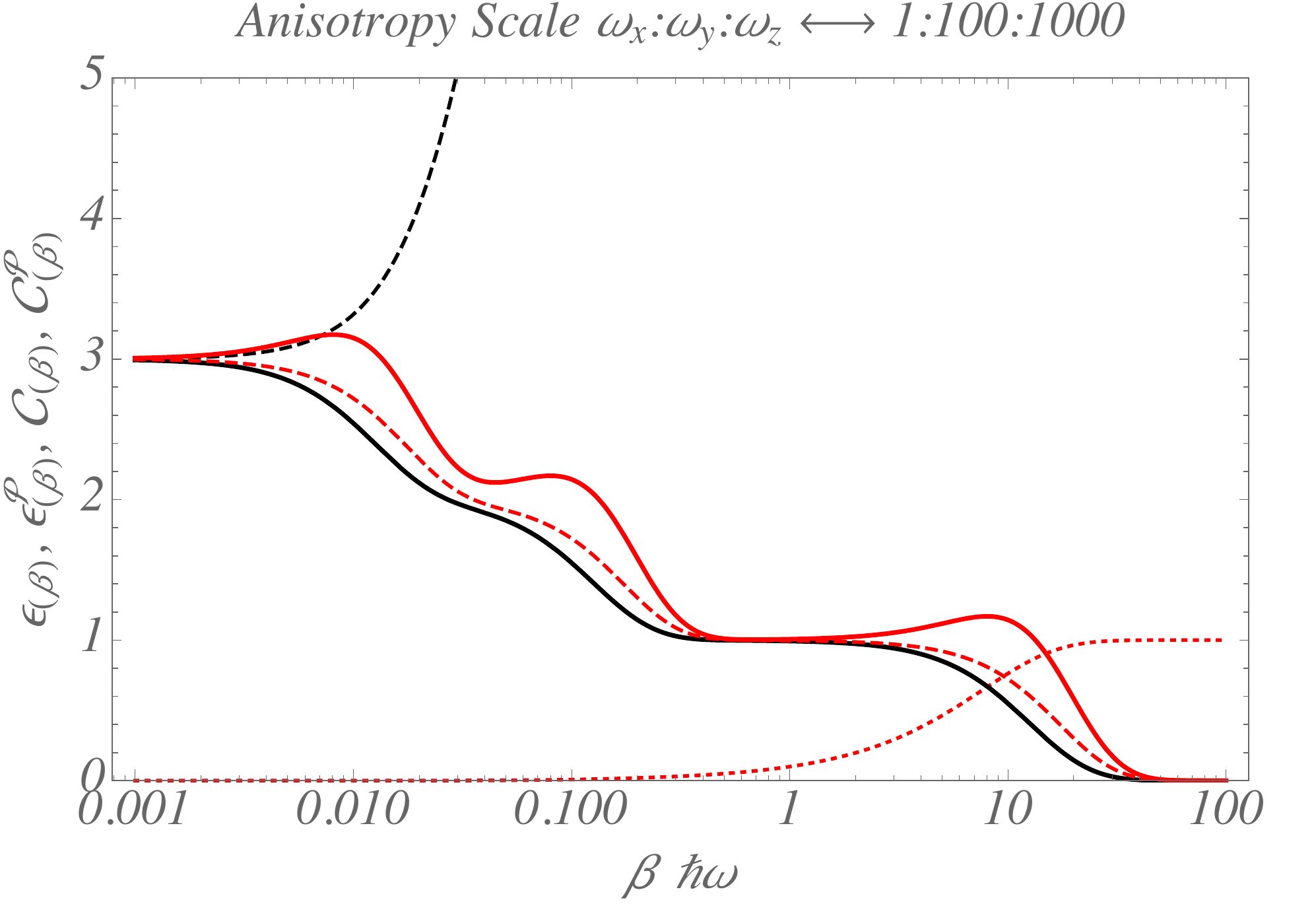}
\renewcommand{\baselinestretch}{.85}
\caption{\footnotesize{
(Color online) Quantum purity, $\mathcal{P}^{HO}_{3D}$ (dotted red lines), and the storage of information capacity described by $\epsilon^{\mathcal{P}}(\beta)$ (dashed red lines) and $\mathcal{C}^{\mathcal{P}}(\beta)$ (solid red lines)
as function of $\beta$.
The plots are for $\omega_x:\omega_y:\omega_z \leftrightarrow 1/10:1:10$ (first plot) and $\omega_x:\omega_y:\omega_z \leftrightarrow 1/10:1:100$ (second plot). By comparative reasons, the quantum ensemble internal energies, $\epsilon(\beta)$ (dashed black lines) and heat capacities, $\mathcal{C}(\beta)$ (solid black lines), are also described as function of $\beta$.}}
\label{fig01}
\end{figure}

Considering the subadditivity properties from Eq.~(\ref{qpqpqpSA}), the results for the quantum purity and their correspondent storage of information properties are much more evinced for an anisotropic $3D$ version the HO system from (\ref{qua16HO3}), for which one has
\begin{equation}\label{qpqpqp}
\mathcal{P}^{HO}_{3D}(\beta) = \tanh\left(\beta\hbar\omega_x/2\right)\tanh\left(\beta\hbar\omega_y/2\right)\tanh\left(\beta\hbar\omega_z/2\right),\end{equation}
which results into the $\epsilon^{\mathcal{P}}(\beta)$ and $\mathcal{C}^{\mathcal{P}}(\beta)$ quantities depicted in Fig.~\ref{fig01}, where the anisotropy property is introduced through the rate of the angular frequencies, $\omega_x:\omega_y:\omega_z$.
From Fig.~\ref{fig01}, it is possible to notice that the results for $\epsilon^{\mathcal{P}}(\beta)$ (dashed red lines) follow a pattern similar to that one of the ensemble heat capacity, $\mathcal{C}(\beta)$ (solid black lines), as well as the storage of information capacity, $\mathcal{C}^{\mathcal{P}}(\beta)$ (solid red lines), let the storage plateaus much more evinced.

Turning back to the generalized interpretation of the quantifiers defined by Eqs.~(\ref{qua16HObb}) and (\ref{qua16HO}), as it can be depicted in both plots from Fig.~\ref{fig01}, the storage of information capacity resumed by $\epsilon^{\mathcal{P}}$ and $\mathcal{C}^{\mathcal{P}}$ -- as they respectively correspond to first and second derivatives of (the logarithm of) $\mathcal{P}$ -- provide more detailed information for the quantum purity variation pattern with respect to the temperature, which sometimes is not evinced from purity profiles by themselves. $\epsilon^{\mathcal{P}}$ and $\mathcal{C}^{\mathcal{P}}$ exhibit similar profiles which however are not identical to those ones of the heat capacity, $\mathcal{C}$. Therefore, they correspond to a set of complementary thermodynamic averaged quantities which quantitatively describe how the variation of the level of quantum decoherence measured by $\mathcal{P}$ evolves in terms of the temperature dependent parameter, $\beta$. Depending on the analytical structure of the partition function, $\mathcal{Z}(\beta)$, without the tools here introduced, the level of decoherence measured by the quantum purity can not be straightforwardly obtained from the energy related thermodynamic variables.

\paragraph*{Singular Oscillator --} An extension of the above results can be exemplified by the {\em singular oscillator} (SO) system driven by the dimensionless Hamiltonian equation \cite{NovoPaper}, 
\begin{equation}\label{qua16}
\mathcal{H}^{SO} \phi^{\alpha}_n(x) = \frac{1}{2}\left\{k^2+ x^{2} + \frac{4 \alpha^2 -1}{4 x^{2}}-2\alpha \right\} \phi^{\alpha}_n(x) = (2n + 1)\,\phi^{\alpha}_n({x}),
\end{equation}
from which the thermodynamic ensemble of quantum states is described in terms of the Wigner distribution written as
\begin{eqnarray}
\label{W222BSO}
\mathcal{W}_\Omega^{\alpha}(x, \, k;\,\beta) &=&\frac{\exp(\beta\hbar\omega)}{\mathcal{Z}(\beta)}\sum_{n=0}^{\infty} \mathcal{W}_n^{\alpha}(x, \, k)\, \exp(-2n\,\beta\hbar\omega),\end{eqnarray}
where $\mathcal{W}_n^{\alpha}$ are the Wigner function contributions obtained from the Weyl transform of the eigenvectors $\phi^{\alpha}_n(x)$ \cite{NovoPaper}, which leads to
\small\begin{eqnarray}\label{geure}
\mathcal{W}_\Omega^{\alpha}(x, \, k;\,\beta) &=& \frac{2 \exp(-{\beta\hbar\omega})}{\mathcal{Z}(\beta)\,\pi}
\int^{+x}_{-x}
\hspace{-.3cm}dy\,\exp\left(2\,i\, k\,y\right)\,\exp\left[-(x^2+y^2)\right]\,(x^2-y^2)^{\frac{1}{2}+\alpha}\times\\
&&\qquad\qquad\qquad
\sum_{n=0}^{\infty}\bigg{\{}\exp(-2n\,\beta\hbar\omega)\frac{n!}{\Gamma(\alpha+n+1)}L_n^{\alpha}\left((x+y)^2\right)\,L_n^{\alpha}\left((x-y)^2\right)\bigg{\}},\quad\quad\quad
\nonumber
\end{eqnarray}\normalsize
where $L_n^{\alpha}$ are the {\em associated Laguerre polynomials}.
In this case, some straightforward mathematical manipulations \cite{Gradshteyn} over Eq.~(\ref{geure}) results into 
\begin{eqnarray}
\label{W222C}
\lefteqn{\sum_{n=0}^{\infty}\bigg{\{}\exp(-2n\,\beta\hbar\omega)\frac{n!}{\Gamma(\alpha+n+1)}L_n^{\alpha}\left((x+y)^2\right)\,L_n^{\alpha}\left((x-y)^2\right)\bigg{\}} =} \\&&\quad\quad\quad\quad\quad\quad\quad\quad\quad\quad\quad\quad\frac{(x^2-y^2)^{-\alpha}}{(1-\lambda)\lambda^{\frac{\alpha}{2}}}
\exp\left[-\frac{2\lambda}{1-\lambda}(x^2+y^2)\right] \mathcal{I}_{\alpha} \left(\frac{2\lambda^{\frac{1}{2}}}{1-\lambda}(x^2-y^2)\right),\quad\quad\quad\nonumber
\nonumber
\end{eqnarray}\normalsize
with $\lambda = \exp(-{2\beta\hbar\omega})$, where $\mathcal{I}_{\alpha}$ is the {\em modified Bessel function of the first kind}, and which can be substituted into Eq.~\eqref{geure} as to give
\small\begin{eqnarray}\label{finalwigner}
\mathcal{W}_\Omega^{\alpha}(x, \, k;\,\beta) &=& \frac{\exp({\alpha\beta\hbar\omega})}{\sinh(\beta\hbar\omega)\mathcal{Z}(\beta)\,\pi}\times\\&&\qquad
\int^{+x}_{-x}
\hspace{-.3cm}dy\,\exp\left(2\,i\, k\,y\right)\,(x^2-y^2)^{\frac{1}{2}}
\exp\left[-\coth(\beta\hbar\omega)(x^2+y^2)\right]\,\mathcal{I}_{\alpha} \left(\frac{x^2-y^2}{\sinh(\beta\hbar\omega)}\right).\nonumber
\end{eqnarray}\normalsize

From the standard statistical definitions (cf. Eq.~(\ref{stand})), and for Hamiltonian eigenvalues described by Eq.~(\ref{qua16}), it is possible to verify that the expressions for the partition function as well as for the quantum purity of the SO do not depend on the anharmonic distortion driven by $\alpha$. In fact, they reproduce exactly the same thermodynamic phenomenology of the HO when $\omega$ is replaced by $2\omega$, which can be expressed by
$\mathcal{Z}^{HO}(2\beta) = \mathcal{Z}^{SO}(\beta)$ and $\mathcal{P}^{HO}(2\beta) = \mathcal{P}^{SO}(\beta)$.
In this sense, partition functions and quantum purities are useless in distinguishing one system from each other.
Otherwise, one can recover the quantum projector definition from Eq.~(\ref{qp}) as to compute $\mathcal{F}^{ab}(2\beta,\,\beta)$ for $a\equiv HO$ and $b \equiv SO$, for the Wigner functions respectively obtained from Eqs.(\ref{W222BHO}) and (\ref{finalwigner}).

\begin{figure}[t!]
\vspace{-.5cm}\includegraphics[scale=0.4]{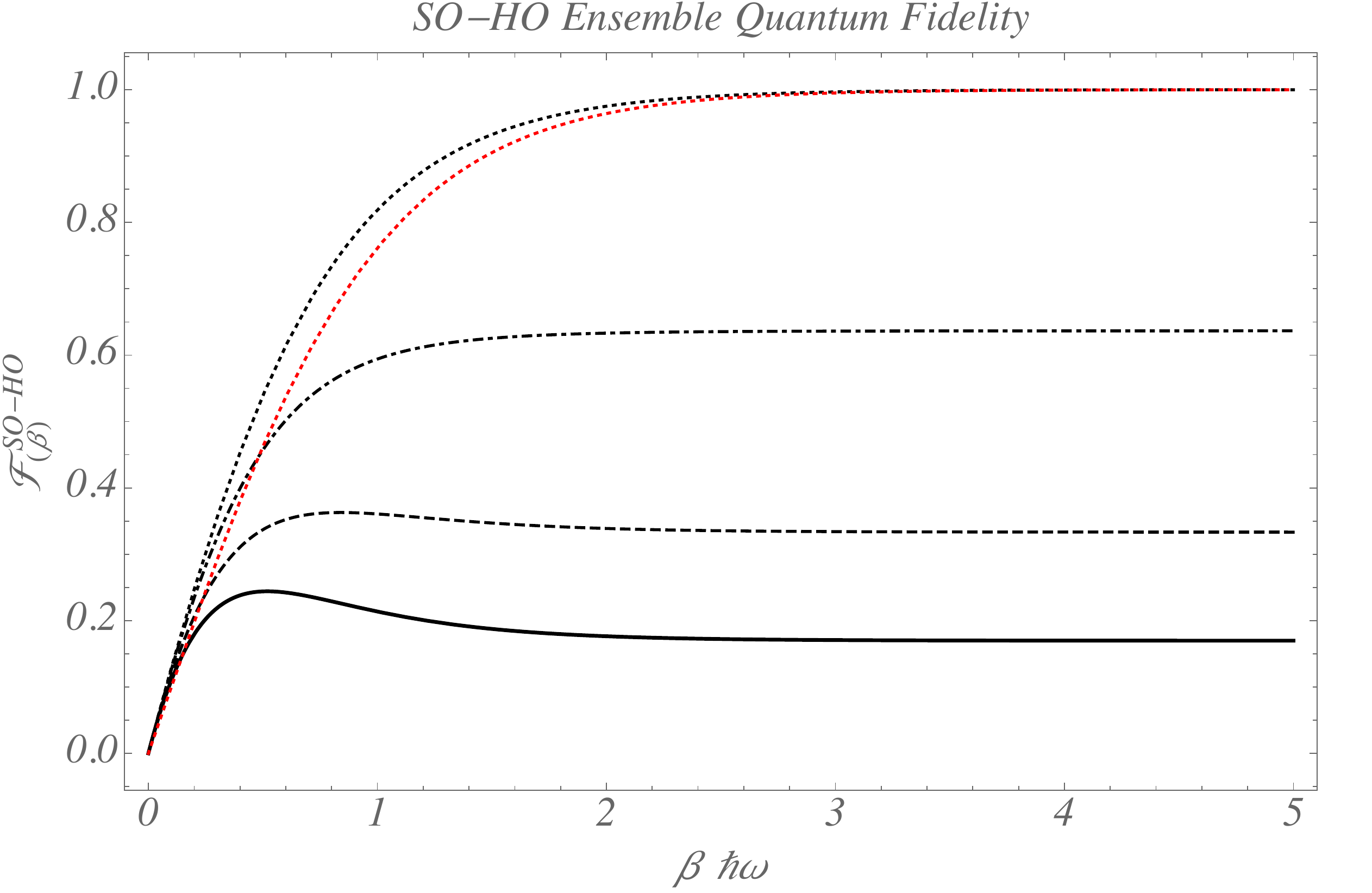}
\renewcommand{\baselinestretch}{.85}
\caption{\footnotesize{
(Color online) Square of quantum fidelity, $\mathcal{F}^{HO-SO}(2\beta,\,\beta)$ between SO and HO thermodynamic ensembles as function of $\beta$. The plots are for $\alpha = -1/2$ (dotted black line), $\alpha = 1/2$ (dot-dashed black line), $\alpha = 3/2$ (dashed black line), and $\alpha = 5/2$ (solid black line).
The asymptotic value of $\mathcal{F}^{HO-SO}$ for $\alpha = -1/2$ recovers the unitary quantum purity results described by $\mathcal{P}^{HO}(2\beta)$ (dashed red line).
}}
\label{fig02}
\end{figure}

Given the {\em quasi}-gaussian profile from (\ref{finalwigner}) and the typical gaussian pattern from (\ref{W222BHO}), the numerically obtained results for $\mathcal{F}^{ab}(2\beta,\,\beta)$ depicted in Fig.~\ref{fig02} can be read as the (square of) the quantum fidelity between SO and HO thermodynamic ensembles, which evinces the role of the parameter $\alpha$ in the anharmonic contributions.

For $\alpha = -1/2$, one approximately recovers the result for the $HO$ quantum purity, with a short distortion due to the $-2\alpha$ factor contribution from the Hamiltonian Eq.~(\ref{qua16}), which is suppressed in the limit of vanishing temperatures (pure state limit, for $\beta \to \infty$).

\paragraph*{Infinite Box and Quantum Rotors --} To summarize, two other typical quantum ensemble systems that can be discussed in the above context are described in terms of the self-energies of the infinite box (1D) and of the quantum rotor (3D).
The semi-analytic expression for their partition functions are respectively given by
\begin{eqnarray}
\mathcal{Z}^{Box}(\beta) &=& \sum_{n=0}^{\infty}\exp\left(-\beta \pi^2 \hbar^2 n^2/(2 m a^2)\right)\nonumber\\ &\approx&\int_0^{\infty}dn \,\exp\left(-\beta \pi^2 \hbar^2 n^2/(2 m a^2)\right) = a\left(\frac{m}{\beta\hbar}\right)^{1/2},
\end{eqnarray}
where $m$ is the mass parameter, and $a$ is the box width parameter, and
\begin{eqnarray}
\mathcal{Z}^{Rot}(\beta) &=&
\sum_{\ell=0}^{\infty}(2\ell+1)\exp\left(-\ell(\ell+1) \beta \hbar^2 /(2 I)\right)\nonumber\\ &\approx&
\int_0^{\infty}d\ell \,2\ell+1)\exp\left(-\ell(\ell+1) \beta \hbar^2 /(2 I)\right) 
\nonumber\\ &=& -\frac{2I}{\beta \hbar^2}\exp\left(-\ell(\ell+1) \beta \hbar^2 /(2 I)\right)\bigg{\vert}_{\ell=0}^{\ell=\infty}=\frac{2I}{\beta \hbar^2},
\end{eqnarray}
where $I$ is the inertia momentum parameter.
Before discussing the above results, it is convenient to notice from Eq.~(\ref{qpqpqp}) that, for a thermodynamic ensemble with the partition function proportional to a power of $\beta$ (or $T$), i.e. $\mathcal{Z}(\beta) \propto \beta^m$, one has $\mathcal{P}(\beta) \propto  1/\mathcal{Z}(\beta)$ which, according to the definitions from Eqs.~(\ref{qua16HObb}) and (\ref{qua16HO}), straightforwardly leads to
\begin{equation}\label{qua16HObbb}
\epsilon^{\mathcal{P}}(\beta) =\epsilon(\beta)=-
\beta \frac{\partial}{\partial \beta} \ln\left(\mathcal{Z}(\beta)\right),
\end{equation}
and
\begin{equation}\label{qua16HOb}
\mathcal{C}^{\mathcal{P}}(\beta) =\mathcal{C}(\beta) =  \beta^2 \frac{\partial^2}{\partial \beta^2} \ln\left(\mathcal{Z}(\beta)\right).
\end{equation}

It means that, for the above related thermodynamic ensembles, the storage of information capacities given in terms of $\epsilon^{\mathcal{P}}$ and $\mathcal{C}^{\mathcal{P}}$ exhibit the same respective pattern described by the internal energy and by the heat capacity obtained from $\mathcal{Z}(\beta)$. Besides a natural consistency interpretation, it supports the relevance of the novel quantifiers here introduced, in the sense that, as noticed from the preliminary results, they introduce a description of complementary information aspects related to the quantum nature of thermodynamic ensembles.

\paragraph*{Brief Conclusion --} In this report, from an strict theoretical perspective, it has been demonstrated that the extension of the definition of phase-space quantum states to thermodynamic ensembles provides general expressions for the quantum purity quantifier and for more generalized phase-space quantum projectors for thermalized quantum systems. The results were explicitly obtained in terms of a straightforward correspondence with thermodynamic ensemble partition functions, $\mathcal{Z}(\beta)$, which supports their statistical interpretation, namely when it is related to subadditivity properties. 
Besides the quantum mechanical support given by the Weyl-Wigner framework, and the expected correspondence with the usual thermodynamic internal energy and heat capacity quantifiers, the information quantifiers here identified introduce novel attributes to the partition functions.
Our results suggest that the so-called storage of information capacities obtained from the quantum projection tools here introduced stir up the comprehension of complementary aspects related to the quantum nature of thermodynamic ensembles. 

As a concluding remark, it is worth mentioning that the Weyl-Wigner representation of the thermodynamic ensembles considered here naturally circumvents the negative ({\em quasi}) probability misunderstandings frequently pointed out in phase-space probability distribution analysis. In this context, an exhausting list of derived frameworks, which includes Husimi $Q$ \cite{Husimi,Ballentine} and Glauber-Sudarshan \cite{Glauber,Sudarshan} phase-space representations, as well as optical tomographic quantum mechanics \cite{Amosov,Radon,Mancini}, can also deserve sharper investigations in the context of thermodynamics.
In this sense, our results can be applied into a broad context of physical problems, from the improvement of quantum information protocols related, for instance, to the computation of loss of information in the thermodynamic cycles of quantum heat engines \cite{Novo01,Novo02,Novo03,Novo04,Novo00}, to the investigation of quantum to classical transitions at cosmological scenarios \cite{JCAP18}.

{\em Acknowledgments} -- This work was supported by the Brazilian agencies FAPESP (grant 2018/03960-9) and CNPq (grant 301000/2019-0).\\

\end{document}